\documentclass[aps,prl,twocolumn,superscriptaddress,longbibliography]{revtex4-1}
\usepackage{mathrsfs,braket}
\usepackage{amssymb, amsbsy, amsmath, latexsym, dsfont, array, layout,
graphicx,mathrsfs,color,ulem,bm}

\newcommand{\one}{\mathds{1}}

\begin{document}

\title{Topology with broken parity-time symmetry}
\author{Lei Xiao}
\affiliation{Beijing Computational Science Research Center, Beijing 100084, China}
\affiliation{Department of Physics, Southeast University, Nanjing 211189, China}
\author{Xingze Qiu}
\affiliation{Key Laboratory of Quantum Information, University of Science and Technology of China, CAS, Hefei 230026, China}
\affiliation{Synergetic Innovation Center in Quantum Information and Quantum Physics, University of Science and Technology of China, CAS, Hefei 230026, China}
\author{Kunkun Wang}
\affiliation{Beijing Computational Science Research Center, Beijing 100084, China}
\affiliation{Department of Physics, Southeast University, Nanjing 211189, China}
\author{Barry C. Sanders}
\affiliation{Shanghai Branch, National Laboratory for Physical Sciences at Microscale, University of Science and Technology of China, Shanghai 201315, China}
\affiliation{Institute for Quantum Science and Technology, University of Calgary, Alberta T2N 1N4, Canada}
\affiliation{Program in Quantum Information Science, Canadian Institute for Advanced Research, Toronto, Ontario M5G 1Z8, Canada}
\author{Wei Yi}\email{wyiz@ustc.edu.cn}
\affiliation{Key Laboratory of Quantum Information, University of Science and Technology of China, CAS, Hefei 230026, China}
\affiliation{Synergetic Innovation Center in Quantum Information and Quantum Physics, University of Science and Technology of China, CAS, Hefei 230026, China}
\author{Peng Xue}\email{gnep.eux@gmail.com}
\affiliation{Beijing Computational Science Research Center, Beijing 100084, China}

\begin{abstract}
Topological edge states arise in parity-time ($\mathcal{PT}$)-symmetric non-unitary quantum dynamics but have so far only been discussed in the $\mathcal{PT}$-symmetry-unbroken regime. Here we report the experimental detection of robust topological edge states in one-dimensional photonic quantum walks with spontaneously broken $\mathcal{PT}$ symmetry, thus establishing the existence of topological phenomena therein. We theoretically prove and experimentally confirm that the global Berry phase in non-unitary quantum-walk dynamics unambiguously defines topological invariants of the system in both the $\mathcal{PT}$-symmetry-unbroken and broken regimes. As topological edge states exist in both $\mathcal{PT}$ unbroken and broken regimes, we reveal that topological phenomena are not driven by $\mathcal{PT}$ symmetry.
\end{abstract}

\maketitle

Topological phases exhibit remarkable properties and challenge our understanding of phases and phase transitions~\cite{HKrmp10,QZrmp11}.
Instead of local order parameters, topological phases are characterized by non-local topological invariants, which dictate the existence and number of topological edge states at an interface through the bulk-boundary correspondence~\cite{Ryu10,Kane10}. Photonic quantum walks (QWs)~\cite{BMK+99,DSB+05,PLM+10,SCP+11,JDL+13,COR+13} offer a versatile platform on which
topological phenomena can be simulated and studied in quantum dynamics~\cite{CSPRA,KB+12,Cardano2016,Cardano2017,BNE+17,RAA17,WXQ+18}. Due to the ease of introducing loss, photonic QWs allow exploration of topological phenomena in the context of non-unitary dynamics~\cite{Zeunerprl,pxprl,PTsymm2}. Recent experimental observations of topological edge states in parity-time ($\mathcal{PT}$)-symmetric systems have stimulated effort in clarifying the relation between topology and $\mathcal{PT}$ symmetry~\cite{PTsymm2,Weimannnm,PBKMS15} as symmetry and topology are two of the conceptual pillars that underlie our understanding of the physics.

However, topological invariants and edge states have so far only been discussed in the $\mathcal{PT}$-symmetry-unbroken regime~\cite{PTsymm2,Weimannnm,PBKMS15,schomerus2013,Harari,KMKO16}, where eigenenergies of the $\mathcal{PT}$-symmetric non-Hermitian Hamiltonian are entirely real~\cite{BB98,BBJ02,B07,KFM08,R+10,P+14,Ch+14,R+12}. In the $\mathcal{PT}$-symmetry-broken regime, where eigenenergies become complex, commonly used topological invariants such as the winding number and the Zak phase become ill-defined~\cite{PTsymm2,Weimannnm,PBKMS15,schomerus2013,Harari,KMKO16}, and thus a series of questions are raised naturally. Will the edge states be observed in $\mathcal{PT}$-broken regimes? If yes, how to explain as the topological invariants are ill-defined in the commonly used definition? Can we define
topological invariants uniformly working well in both the $\mathcal{PT}$-symmetric-unbroken and -broken regimes? Is there any interplay between topology and $\mathcal{PT}$ symmetry if there exist the topological protected edge states in both $\mathcal{PT}$-symmetric-unbroken and -broken regimes?

In this work, we answer these questions. We observe robust topological edge states in one-dimensional photonic QWs with spontaneously broken $\mathcal{PT}$ symmetry, thus establishing the existence of topological phenomena therein. We theoretically prove and experimentally confirm that the global Berry phase in non-unitary QW dynamics gives rise to well-defined topological invariants in both the $\mathcal{PT}$-symmetry-unbroken and broken regimes, which are responsible for the emergence of topological edge states. Our results establish a unified framework for characterizing topological phenomena in non-unitary QW dynamics, and provide a solid foundation for future investigations of topological phenomena in $\mathcal{PT}$-symmetric non-Hermitian systems in general. As topological edge states are $\mathcal{PT}$-symmetry broken, we reveal for the first time that topological phenomena in $\mathcal{PT}$-symmetric non-Hermitian systems are not driven by $\mathcal{PT}$ symmetry.

\begin{figure*}
\includegraphics[width=\textwidth]{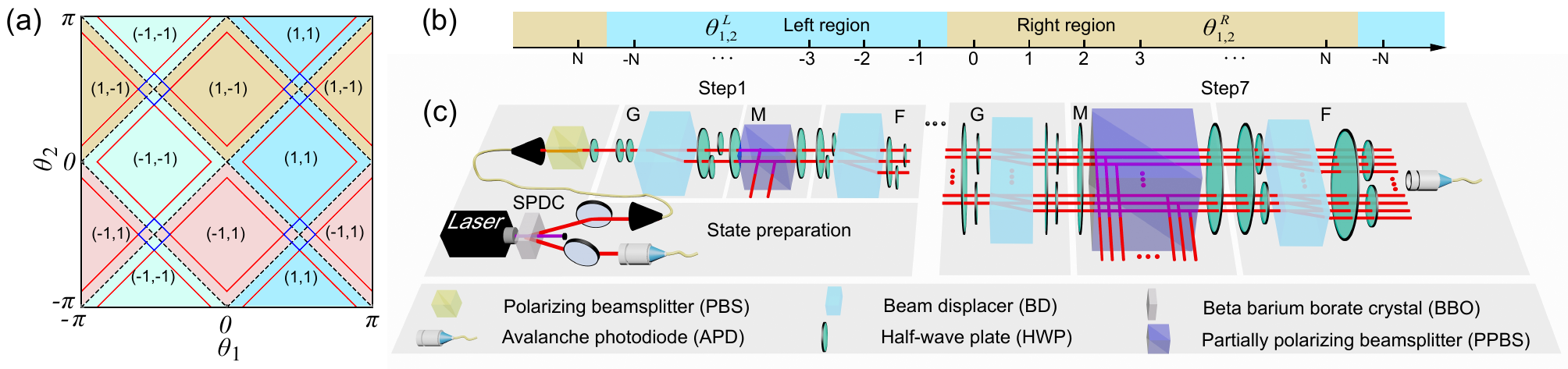}
\caption{(a) Phase diagram for homogenous QWs governed by $\tilde{U}'$, with the coin parameters $(\theta_1,\theta_2)$ and the corresponding topological numbers $(\nu_0,\nu_\pi)$. Dashed black lines represent topological phase boundaries. Solid red lines represent boundaries between $\mathcal{PT}$-symmetry-unbroken and broken regimes, with $\mathcal{PT}$-symmetry-broken regimes lying inbetween the red lines near topological phase boundaries.
Solid blue squares represent regimes with completely broken $\mathcal{PT}$ symmetry, where the eigenspectra are purely imaginary~\cite{supp}.
(b) Left ($x<0$) and right ($x \geq 0$) regions for the $\mathcal{PT}$-symmetric QW. (c) Experimental setup for $\mathcal{PT}$-symmetric QWs with alternating losses. The photon pair is created via spontaneous parametric downconversion. One photon serves as a trigger. The other is projected into the polarization state $\ket{\pm}$ (or $(\ket{+}+i\ket{-})/\sqrt{2}$) and then proceeds through the quantum-walk interferometric network. Finally, the photon is detected by avalanche photodiode (APD), in coincidence with the trigger one. Photon counts give measured probabilities after correcting for relative efficiencies of the different APDs.}
\label{fig:fig1}
\end{figure*}

{\bf Results}

{\it $\mathcal{PT}$-symmetric QWs.} We consider $\mathcal{PT}$-symmetric non-unitary QWs corresponding to an alternating gain-loss scheme, where the evolution in each time step is governed by the Floquet operator
\begin{equation}
\tilde{U}'=F\tilde{M}G,
\label{eq:floquet}
\end{equation}
with
\begin{align}
&F=R\left[\frac{\theta_1(x)}{2}\right]SR
\left[\frac{\theta_2(x)}{2}\right],\\ \nonumber
&G=R\left[\frac{\theta_2(x)}{2}\right]SR\left[\frac{\theta_1(x)}{2}\right],\\ \nonumber
& \tilde{M}=\gamma\left[\one_\text{w}\otimes\left(\ket{+}\bra{+}+\sqrt{1-p}\ket{-}\bra{-}\right)\right],\quad 0<p\leqslant 1.
\end{align}
Here the QW is on a one-dimensional integer lattice $\mathbb{L}$ on a circle, with site index $-N\leq x \leq N$ and $N$ being the largest positive site index.
The conditional-shift operator $S$ moves the walker in the two orthogonal coin states $|0\rangle$ and $|1\rangle$, respectively, to the left and right by one lattice site. The position-dependent coin operator $R\left[\theta(x)\right]$ rotates the coin state by $\theta(x)$ about the $y$-axis. $\tilde{M}$ is the gain-loss operator by which non-unitarity is enforced with $\ket{\pm}=(\ket{0}\pm\ket{1})/\sqrt{2}$, $\gamma=(1-p)^{-\frac{1}{4}}$, and $\one_\text{w}=\sum_{\mathbb{L}}\ket{x}\bra{x}$. Under $\tilde{M}$, states in $|\pm\rangle$ are amplified/suppressed by $\gamma^{\pm1}$ in each step.

The non-unitary operator $\tilde{U}'$ is $\mathcal{PT}$-symmetric as long as the coin parameters satisfy $\theta_{1,2}(x)=\theta_{1,2}(N-x)$. Under the periodic boundary condition, the symmetry operator is $\mathcal{PT}=\sum_{x\in \mathbb{L}}|x\rangle\langle N-x|\otimes \sigma_z\mathcal{K}$, with $\mathcal{PT}\tilde{U}'\left(\mathcal{PT}\right)^{-1}=\tilde{U}'^{-1}$, where $\mathcal{K}$ is complex conjugation. In this paper, $\tilde{U}'$ is different from the $\mathcal{PT}$-symmetric case~\cite{PTsymm2} for a simpler gain-loss mechanism which increases measurement efficiency and accuracy and is easier to implement. The current setup also allows us to focus on edge states near $x=0$, when $N$ is much larger than the number of time steps of QW dynamics initialized at $x=0$.
Furthermore, $\tilde{U}'$ is different from the Floquet operator in~\cite{pxprl}, which has no explicit $\mathcal{PT}$ symmetry. The new experimental design here is crucial for establishing topology in the $\mathcal{PT}$-symmetry-broken regime and further revealing that topological phenomena in $\mathcal{PT}$-symmetric non-Hermitian systems are not driven by $\mathcal{PT}$ symmetry.

We define the eigenvalue $\lambda$ and quasienergy $\epsilon$ of $\tilde{U}'$ through
\begin{align}
\tilde{U}'|\psi_{\lambda}\rangle=\lambda|\psi_{\lambda}\rangle,
\quad \lambda= e^{-i\epsilon}.
\end{align}
In the $\mathcal{PT}$-symmetry-unbroken regime, $\{\epsilon\}$ are real and $|\lambda|=1$. Otherwise, when the system has spontaneously broken $\mathcal{PT}$ symmetry,
some $\epsilon$ become complex as $|\lambda|\neq 1$. We further divide the $\mathcal{PT}$-symmetry-broken regime into partially-broken and completely-broken regimes, with $\{\epsilon\}$ being purely imaginary in the latter.
The boundaries between these regimes are shown by the solid lines in Fig.~\ref{fig:fig1}(a).

\begin{figure*}
  \includegraphics[width=0.85\textwidth]{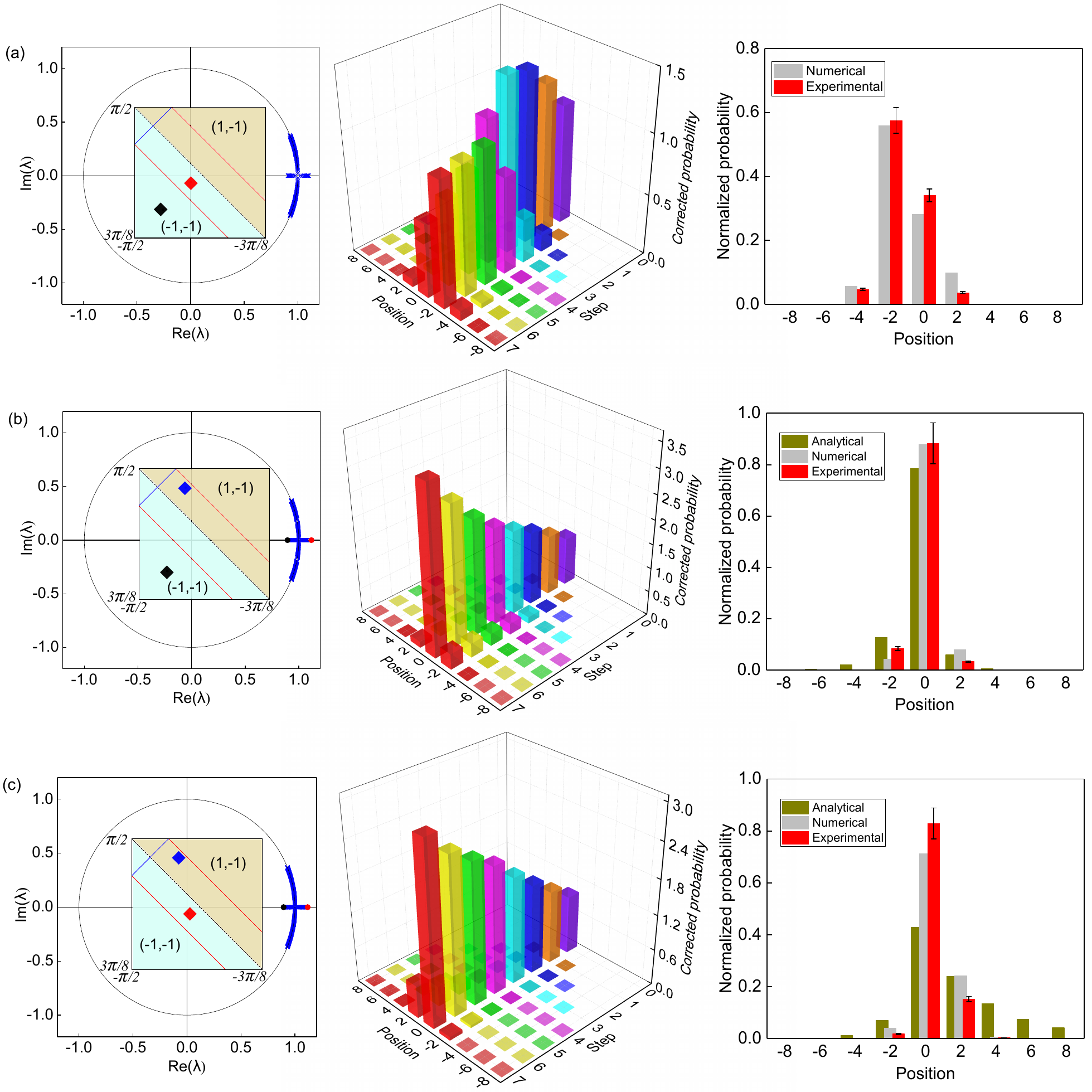}
   \caption{Experimental observation of topological edge states in the $\mathcal{PT}$-symmetry-broken regime.
   (a)-(b) We fix the coin parameters $(\theta^\text{R}_1,\theta^\text{R}_2)=(-7\pi/16-\xi,7\pi/16-\xi)$ in the right region, and vary those in the left region, with (a) $(\theta^{\text{L}}_1,\theta_2^{\text{L}})=(-7\pi/16-\xi/4,7\pi/16-\xi/4)$ and (b) $(-15\pi/32+3\xi/8,15\pi/32+3\xi/8)$.
   (c) We set $(\theta^\text{R}_1,\theta^\text{R}_2)=(-7\pi/16-\xi/4,7\pi/16-\xi/4)$ and $(\theta^{\text{L}}_1,\theta_2^{\text{L}})=(-15\pi/32+3\xi/8,15\pi/32+3\xi/8)$. We fix $\xi=0.1113$ and $p=9/25$ here.
Inset: Phase diagram, with symbols indicating the coin parameters and the corresponding topological numbers for each experimental case.
Left column: eigenvalues $\lambda$ in the complex plane.
Central column: measured corrected probability distributions up to seven steps. Right column: comparison between the measured, numerically calculated, and analytically calculated normalized probability distributions at the seventh step. Experimental errors are due to photon-counting statistics and represent the corresponding standard deviations.
   }
\label{fig:fig3}
\end{figure*}

\begin{figure}
\includegraphics[width=0.5\textwidth]{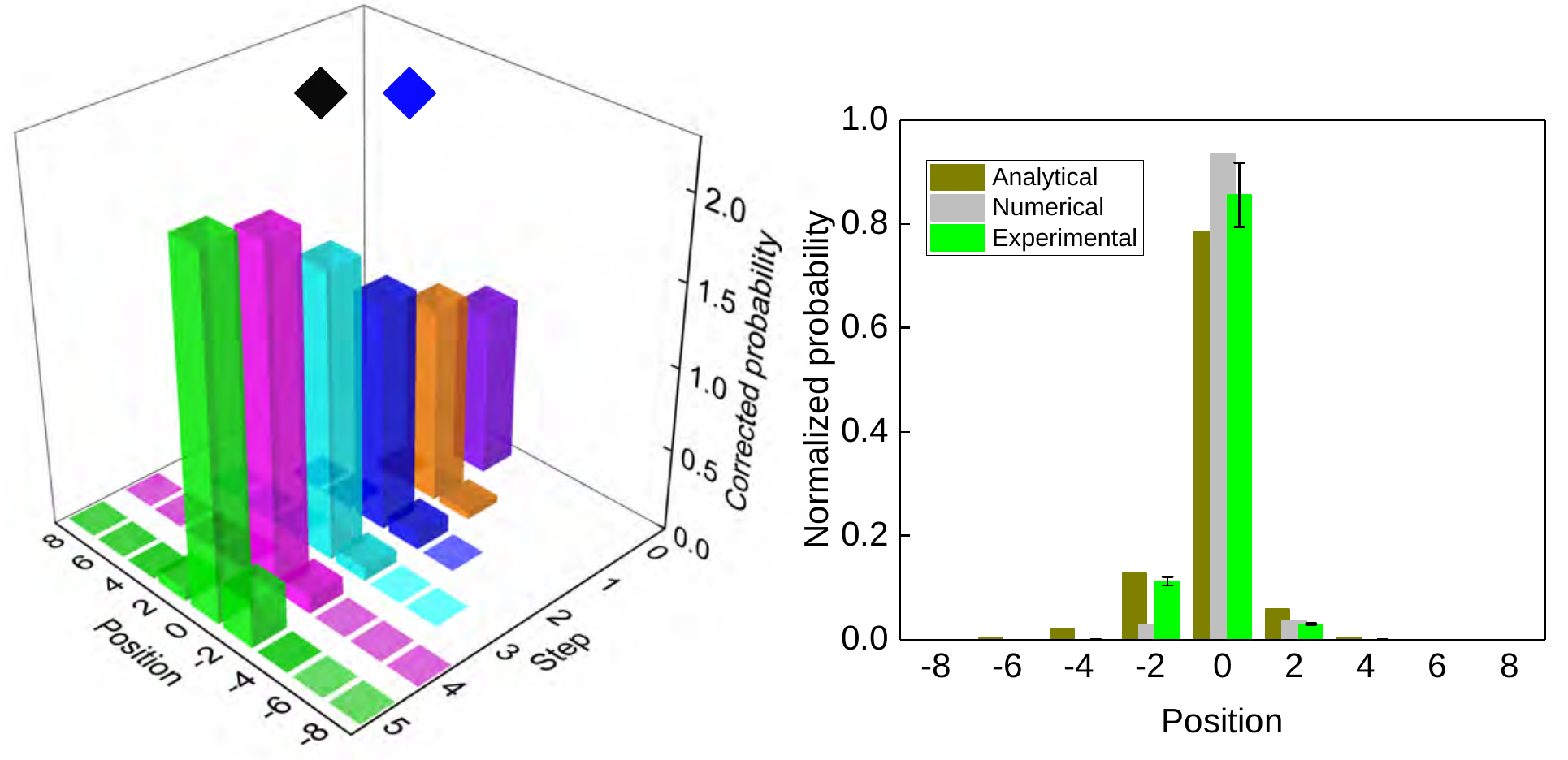}
\caption{Robustness of edge states against static disorder. Probability distributions of five-step QWs with the initial state $\ket{0}\otimes\ket{+}$. The coin parameters are $(\langle\theta^\text{R}_1\rangle,\langle\theta^\text{R}_2\rangle)=(-7\pi/16-\xi,7\pi/16-\xi)$ and $(\langle\theta^\text{L}_1\rangle,\langle\theta^\text{L}_2\rangle)=(-15\pi/32+3\xi/8,15\pi/32+3\xi/8)$, as same as in Fig.~\ref{fig:fig3}(b).
Parameters $p$ and $\xi$ are the same as those in Fig.~\ref{fig:fig3} too. The disordered rotation angles are given by $\theta_{1,2}+\delta\theta$, where $\delta\theta$ is unique for each position and is independent of time and chosen from the intervals $\left[-\xi/4,\xi/4\right]$. Left column: measured corrected probability up to five steps. Right column: comparison between the measured and numerically calculated normalized probability distribution at the fifth step, as well as that calculated from the analytical edge-state wave functions. 
}
\label{fig:fig4}
\end{figure}

{\bf Topological invariant through global Berry phase.} In Eq.~(\ref{eq:floquet}), with $p=0$, $\tilde{U}'$ is unitary and recovers chiral symmetries. Thereby, the system with $p=0$ possesses non-trivial topological phases driven by chiral symmetry. However, for $0<p<1$, chiral symmetry is broken. $\tilde{U}'$ becomes non-unitary and gives rise to Floquet topological phases (FTPs) in the dynamics, ensured by pseudo-anti-unitarity~\cite{ESHK11}, with $\eta \tilde{U}'^{\dag}\eta=\tilde{U}'$ and $\eta=\sum_{x\in \mathbb{L}}|x\rangle\langle x|\otimes\sigma_x$. Here $\sigma_j$ ($j=x,y,z$) are Pauli matrices.

We characterize FTPs associated with the non-unitary operator $\tilde{U}'$ using topological numbers given by the global Berry phase. For the convenience of calculation, we apply a unitary transformation
$W' = V\tilde{U}'V^\dag$, with $V=e^{i\frac{\pi/2}{2}\sigma_y}$. Topological properties of $\tilde{U}'$ is not changed under the unitary transformation.

As $W'$ is non-unitary, we define the left and right eigenstates, respectively, as $\langle \chi_{\pm}|$ and $|\psi_{\pm}\rangle$, with $W'^{\dag}|\chi_\pm\rangle=\lambda_\pm^* |\chi_\pm\rangle$ and $W'|\psi_\pm\rangle=\lambda_\pm|\psi_\pm\rangle$. The left and right eigenstates satisfy the orthonormal conditions $\langle\chi_{\pm}|\psi_{\pm}\rangle = 1$, $\langle\chi_{\pm}|\psi_{\mp}\rangle = 0$. The global Berry phase $\varphi_{\text{B}}$ is then defined as
\begin{align}
\varphi_{\text{B}}&=\varphi_{Z+}+\varphi_{Z-},\label{eqn:gb}\\
\varphi_{Z\pm}&=-i\oint \text{d}k\frac{\langle\chi_{\pm}|\frac{\text{d}}{\text{d}k}|\psi_{\pm}\rangle}{\langle\chi_{\pm}|\psi_{\pm}\rangle},\label{eqn:gb2}
\end{align}
where $\varphi_{Z\pm}$ are the generalized Zak phases for the two bands ($\mu=\pm$). Compared to the previous definition of topological number from the Zak phase $\varphi_{Z-}/2\pi$ which only works in the $\mathcal{PT}$-unbroken regime, here we define the topological number from the global Berry phase as $\nu'=\varphi_{\text{B}}/2\pi$ uniformly working well in both the $\mathcal{PT}$-symmetric-unbroken and -broken regimes.

The integrand in $\varphi_{Z\pm}$ (\ref{eqn:gb2}) is well-defined throughout the first Brillioune zone in the $\mathcal{PT}$-symmetry-unbroken regime or in the regime where $\mathcal{PT}$ symmetry is completely broken. However, in the $\mathcal{PT}$-symmetry-partially-broken regime, the bulk gap vanishes at $0$ or $\pi$ at discrete momenta. At those points, we no longer have orthonormal conditions between the left and right eigenstates, which makes the integrands for $\varphi_{Z\pm}$ divergent. However, as we show in the Supplemental Materials, at these discrete momenta, the divergence in the integrand for $\varphi_{Z+}$ cancels with that for $\varphi_{Z-}$, such that $\varphi_\text{B}$ (\ref{eqn:gb}) remains well-defined and finite. This allows us to extend the definition of topological invariants to $\mathcal{PT}$-symmetry broken regimes to account for the topological edge states observed at interfaces where coin parameters in at least one of the adjacent bulks are in the $\mathcal{PT}$-symmetry-broken regime. We note that at topological phase boundaries, both $\varphi_\text{B}$ and $\varphi_{Z\pm}$ are ill-defined.

As $\nu'$ does not contain enough information to characterize two topological numbers for two different quasienergies $\text{Re}(\epsilon)=0$ and $\text{Re}(\epsilon)=\pi$, we need to treat other Floquet operator $\tilde{U}''=G\tilde{M}F$ fitting in a different time frame. We then define another winding number $\nu''$ through the global Berry phase of $\tilde{U}''$~\cite{PTsymm2,AO13}. The topological numbers for the edge states at $\text{Re}(\epsilon)=0$ and $\text{Re}(\epsilon)=\pi$ are constructed as $(\nu_0,\nu_{\pi})=(\frac{\nu'-\nu''}{2},\frac{\nu'+\nu''}{2})$.
From numerical calculations, we confirm that the topological numbers $(\nu_0,\nu_{\pi})$ are directly related to the number of topological edge states at a given interface. Specifically, the number of edge states with quasienergy $\text{Re}(\epsilon)=0$ [$\text{Re}(\epsilon)=\pi$] is  equal to the difference of topological numbers $\nu_0$ ($\nu_{\pi}$) on either side of the boundary, as we demonstrate numerically in the Supplemental Materials. As illustrated in Fig.~\ref{fig:fig1}(a), different topological phases are labeled by distinct topological numbers $(\nu_0,\nu_{\pi})$, whereas
$\mathcal{PT}$ symmetry is spontaneously broken in the vicinity of topological phase boundaries.

{\bf Topological edge states.} Topological numbers adopted above are equivalent to winding numbers~\cite{PBKMS15,Weimannnm,schomerus2013,Harari} or generalized Zak phases~\cite{PTsymm2,KMKO16} in the $\mathcal{PT}$-symmetry-unbroken regime. In the $\mathcal{PT}$-symmetry-broken regime, whereas winding numbers and generalized Zak phases become ill-defined, the global Berry phases remain well-defined and yield topological numbers that dictate the number of topological edge states.

To investigate topological edge states, we consider an inhomogeneous configuration, where interfaces exist near $x=0$ and $x=\pm N$ separating the left (L) and right (R) regions with $(\theta^{\text{L},\text{R}}_1,\theta^{\text{L},\text{R}}_2)$ [see Fig.~\ref{fig:fig1}(b)].
Topological edge states can emerge near $x=0$ and $x=\pm N$, with the number of edge states having $\text{Re}(\epsilon)=0$ [$\text{Re} (\epsilon)=\pi$] equal to the difference in the topological number $\nu_0$ ($\nu_{\pi}$) on either side of the boundary. These topological edge states break $\mathcal{PT}$ symmetry, such that their quasienergies $\epsilon$ are complex with $\epsilon=i\ln\gamma$ ($\lambda=\pm\gamma$) or $\epsilon=\pi+i\ln\gamma$ ($\lambda=\pm 1/\gamma$).
We identify states with $\lambda=\pm\gamma$ ($\lambda=\pm 1/\gamma$) as topological edge states, as their probability distribution would be amplified (suppressed) over time. We analytically solve wave functions for topological edge states localized near $x=0$~\cite{supp}, which agrees with the prediction by the bulk-edge correspondence.



{\bf Experimental implementation.} As illustrated in Fig.~\ref{fig:fig1}(c), we use a photonic setup to implement passive $\mathcal{PT}$-symmetric QW of single photons.
The coin states $|0\rangle$ and $|1\rangle$ are respectively encoded in the horizontal $\ket{H}$ and vertical $\ket{V}$ polarizations of photons, whose spatial modes represent the lattice degrees of freedom.
In our experiment, the initial coin state is prepared in either $\ket{\pm}$ or $(\ket{+}+i\ket{-})/\sqrt{2}$, while the walker always starts from $x=0$. The experimentally realized time-evolution operator is actually $U'=FMG$, which differs from $\tilde{U}'$ only by a scaling factor $\gamma$. We implement the coin-rotation operator $R(\theta)$, the shift operator $S$, and the loss operator $M=\tilde{M}/\gamma$ using appropriate combinations of half-wave plates (HWPs), beam displacers (BDs), and partially polarizing beamsplitters (PPBSs). The loss parameter $p$ is fixed at $9/25$, which is achieved using PPBS with a certain polarization-dependent transmissivity.

For the resulting probabilities, we therefore multiply the measured raw probability distribution $P_\text{R}(x,t)$ by a time-dependent scaling factor $\gamma^{2t}$, so that the resulting corrected probability distribution $P_\text{C}(x,t)=\gamma^{2t} P_\text{R}(x,t)$ corresponds to $\mathcal{PT}$-symmetric QWs governed by $\tilde{U}'$ in Eq.~(\ref{eq:floquet}). For future reference, we define the normalized probability at the $t$-th step $P_\text{N}(x,t)=P_\text{R}(x,t)/\sum_x P_\text{R}(x,t)$.

{\bf Confirming topological invariants.} We confirm the validity of topological invariants defined through the global Berry phase by detecting the topological edge states.

We focus on the case in which the coin parameters for the left or right region are chosen in the $\mathcal{PT}$-symmetry-unbroken regime, i.e., at least one of the bulks is $\mathcal{PT}$-symmetry broken. In Fig.~\ref{fig:fig3}(a), both left and right regions belong to the same topological phase with $(\nu_0,\nu_\pi)=(-1,-1)$, whereas the left region is $\mathcal{PT}$-symmetry broken. No edge state is expected. The measured corrected probability near the boundary $x=0$ is not enhanced, and after several steps of evolution the probability is no longer localized at the boundary, suggesting the absence of edge states. In Fig.~\ref{fig:fig3}(b), we change the left region to the $\mathcal{PT}$-symmetry-broken regime with $(\nu_0,\nu_\pi)=(1,-1)$.
In the central column, our experimental results clearly show the enhancement of the corrected probability near $x=0$, which gets amplified in time. In the right column, the measured normalized spatial probability distribution after the seventh step agrees reasonably well with the probability given by analytical edge-state wave functions.
These observations confirm the existence of topological edge states in the presence of $\mathcal{PT}$-symmetry broken bulks, which indicates the robustness of topological phenomena against spontaneous $\mathcal{PT}$-symmetry breaking.

In Fig.~\ref{fig:fig3}(c), the left and right regions belong to FTPs with different topological numbers $(\nu_0,\nu_\pi)=(-1,-1)$ and $(1,-1)$ respectively, and both regions are $\mathcal{PT}$-symmetry broken. Whereas topological edge states can still be identified through the amplified $P_\text{C}(x=0,t)$, the measured normalized probability distribution after the seventh step is not fully converged to the analytical solution. This suggests that it takes more time steps for the QW dynamics to converge into topological edge states in the presence of $\mathcal{PT}$-symmetry-broken bulks. Nevertheless, similar to the second case above, our results confirm the existence of topological edge states in the presence of $\mathcal{PT}$-symmetry broken bulks. In all three cases, the measured normalized spatial probability distributions $P_\text{N}(x,t=7)$ agree well with that calculated from analytical edge-state wave functions.


{\bf Robustness of edge states against disorder.} A key feature of topologically non-trivial systems is the robustness of topological properties against small perturbations. We experimentally confirm the robustness of the topological edge states by introducing static disorder to the coin rotations. The static disorder breaks $\mathcal{PT}$ symmetry, but preserves the pseudo-anti-unitarity of $\tilde{U}'$.


We study the robustness of topological edge states when the left region is in the $\mathcal{PT}$-symmetry-broken regime and the right region is in the $\mathcal{PT}$-symmetry-unbroken regime, respectively. We introduce static disorder to the coin rotations by modulating the setting angles of the corresponding HWPs by a small random amount $\delta\theta\in\left[-\xi/4,\xi/4\right]$ around $\theta^\text{L,R}_{1,2}$. Here $\delta\theta$ is time-independent and unique for each position. We then measure the probabilities of the walker up to five steps. As shown in Fig.~\ref{fig:fig4}, the measured corrected probability at $x=0$ increases with time (left), while the normalized probability after the fifth step converges to that given by the analytical edge-state wave function (right). These observations confirm the robustness of topological edge states against static disorder even in the $\mathcal{PT}$-symmetry-broken regime.

{\bf Discussion}

By confirming the existence of topological properties in the $\mathcal{PT}$-symmetry-broken regimes, our results clarify the relation between non-unitary dynamics, $\mathcal{PT}$ symmetry, and topology in one-dimensional topological systems with pseudo-anti-unitarity. In particular, our topological invariants are also capable of characterizing topological properties in non-unitary dynamics without explicit $\mathcal{PT}$ symmetry~\cite{RL09,RLL16,Zeunerprl,pxprl},
where topological numbers calculated through the global Berry phase are equivalent to generalized winding numbers associated with complex-valued pseudo-spin vectors of Bloch Hamiltonians. This provides a unified description for non-unitary QW dynamics either with or without explicit $\mathcal{PT}$-symmetry, thus enabling two previously separate branches of research to be understood and treated on common grounds. As topological edge states are $\mathcal{PT}$-symmetry broken, we reveal for the first time that topological phenomena in $\mathcal{PT}$-symmetric non-Hermitian systems are not driven by $\mathcal{PT}$ symmetry but actually by pseudo-anti-unitarity. Our work represents a significant step toward a deeper understanding of topological features in $\mathcal{PT}$-symmetric systems.


{\bf Methods}



{\bf Edge-state wave functions.}
As $\tilde{U}'$ has two shift operators $S$ in each time step, eigen wave functions on odd and even lattice sites are decoupled. As we detail in the Supplemental Materials, wave functions of topological edge states are written as
\begin{align}
|\psi_{m}^{\text{o}(\text{e})}(x)\rangle=
\begin{cases}
r^{\text{o}(\text{e})}e^{\kappa_{\text{L}}x}|m\rangle,~x<0,\\
t^{\text{o}(\text{e})}e^{-\kappa_{\text{R}}x}|m\rangle,~x\geq 0,
\end{cases}
\end{align}
where $\psi^{\text{o}(\text{e})}_{m}$ is the wave function on odd (even) sites for bright ($m=\text{b}$) and dark ($m=\text{d}$) edge states, respectively. The spatial decay rates $(\kappa_{\text{L}},\kappa_{\text{R}})$ and the coefficients $[r^{\text{o}(\text{e})},t^{\text{o}(\text{e})}]$ all have analytical forms, which depend on the coin parameters and the edge-state quasienery.
Quasienergies of bright edge states are $\epsilon^{(g)}_{\text{b}}=g+ i\ln\gamma$ ($g=0,\pi$), and those of dark edge states are $\epsilon^{(g)}_{\text{d}}=g- i\ln\gamma$ ($g=0,\pi$). Correspondingly, the two types of edge states evolve in time according to $(\gamma e^{-ig})^t$ and $(\gamma^{-1} e^{ig})^{t}$, respectively. For a bright edge state with quasienergy $\epsilon_{\text{b}}^{(0)}$ [$\epsilon_{\text{b}}^{(\pi)}$], its coin state satisfies $|\text{b}\rangle=|+\rangle$ ($|\text{b}\rangle=|-\rangle$). For a dark edge state with quasienergy $\epsilon_{\text{d}}^{(0)}$ ($\epsilon_{\text{d}}^{(\pi)}$), its coin state satisfies $|\text{d}\rangle=|-\rangle$ ($|\text{d}\rangle=|+\rangle$). Whereas analytical solutions of bright edge states agree well with experimental measurement, wave functions of both types of edge states are consistent with numerical calculations.

{\bf Experimental implementation of $\tilde{U}'$.}
With a single-photon source consisting of a $\beta$-barium-borate (BBO) nonlinear crystal pumped by a CW diode laser, we generate polarization-degenerate photon pairs at $801.6$nm using a type-I spontaneous parametric down-conversion (SPDC) process.
Upon detection of a trigger photon, the signal photon is heralded in the measurement setup. This trigger-signal photon pair is registered by a coincidence count at two APDs with a $\Delta t=3$ns time window. Total coincidence counts are about $10,000$ over a collection time of $2$s.

The coin states $|0\rangle$ and $|1\rangle$ are respectively encoded in the horizontal $\ket{H}$ and vertical $\ket{V}$ polarizations of the heralded single photon, whose spatial modes represent the walker state. After passing through a polarizing beamsplitter (PBS) followed by a HWP, the heralded single photon is projected into an arbitrary initial state and then proceeds through the quantum-walk interferometric network. We implement the coin operator $R(\theta)=\one_\text{w}\otimes \text{e}^{-i\theta\sigma_y}$ (here $\sigma_y=i(-\ket{H}\bra{V}+\ket{V}\bra{H})$ is the standard Pauli operator under the polarization basis) by HWPs with certain setting angles depending on the coin parameters $(\theta_1,\theta_2)$, and the shift operator $S=\sum_x\left(\ket{x-1}\bra{x}\otimes\ket{H}\bra{H}+\ket{x+1}\bra{x}\otimes\ket{V}\bra{V}\right)$ by a BD whose optical axis is cut so that the photons in $\ket{V}$ are directly transmitted and those in $\ket{H}$ undergo a lateral displacement into a neighboring spatial mode, respectively. The loss operator $M$ is implemented by a sandwich-type HWP (at $22.5^\circ$)-PPBS-HWP (at $22.5^\circ$) setup~\cite{pxprl}. Here the transmissivities of PPBS are $(T_\text{H},T_\text{V})=(1,1-p)$ for horizontally and vertically polarized photons, respectively.

We construct the raw probability distribution of the walker $P_\text{R}$ at time $t$ by dividing the number of coincidence measurements at APDs using the total number of photon pairs, after correcting for the relative efficiencies of different APDs. The raw probability is then converted into the corrected probability $P_\text{C}(x,t)=\gamma^{2t}P_\text{R}(x,t)$, which is obtained by multiplying the correction factor $\gamma$ for the corresponding step $t$ and represents the probability corresponding to $\mathcal{PT}$-symmetric QWs governed by $\tilde{U}'$. Whereas, the normalized probability $P_\text{N}(x,t)$ is defined as $P_\text{R}(x,t)/\sum_x P_\text{R}(x,t)$.


\begin{acknowledgments}
\textit{Acknowledgement:--}
We thank Hideaki Obuse for helpful discussions. This work has been supported by the Natural Science Foundation of China (Grant Nos. 11674056, and 11522545) and the Natural Science Foundation of Jiangsu Province (Grant No. BK20160024). WY acknowledges support from the National Key R\&D Program (Grant Nos. 2016YFA0301700,2017YFA0304100). LX and XQ contributed equally to this work.
\end{acknowledgments}

\bibliography{reference}



\clearpage

\begin{widetext}
\renewcommand{\thesection}{\Alph{section}}
\renewcommand{\thefigure}{S\arabic{figure}}
\renewcommand{\thetable}{S\Roman{table}}
\setcounter{figure}{0}
\renewcommand{\theequation}{S\arabic{equation}}
\setcounter{equation}{0}


\section{Supplemental Materials for ``Topology with broken parity-time symmetry''}


\section{Experimental implementation of $\tilde{U}'$}

With a single-photon source consisting of a $\beta$-barium-borate (BBO) nonlinear crystal pumped by a continuous wave diode laser, we generate polarization-degenerate photon pairs at $801.6$nm using a type-I spontaneous parametric down-conversion (SPDC) process.
Upon detection of a trigger photon, the signal photon is heralded in the measurement setup. This trigger-signal photon pair is registered by a coincidence count at two avalanche photodiodes (APDs) with a $\Delta t=3$ns time window. Total coincidence counts are about $10,000$ over a collection time of $2$s.

The coin states $|0\rangle$ and $|1\rangle$ are respectively encoded in the horizontal $\ket{H}$ and vertical $\ket{V}$ polarizations of the heralded single photon, whose spatial modes represent the walker state. After passing through a polarizing beamsplitter (PBS) followed by a half-wave plate (HWP), the heralded single photon is projected into an arbitrary initial state and then proceeds through the quantum-walk interferometric network. We implement the coin operator $R(\theta)=\one_\text{w}\otimes \text{e}^{-i\theta\sigma_y}$ (here $\sigma_y=i(-\ket{H}\bra{V}+\ket{V}\bra{H})$ is the standard Pauli operator under the polarization basis) by HWPs with certain setting angles depending on the coin parameters $(\theta_1,\theta_2)$, and the shift operator $S=\sum_x\left(\ket{x-1}\bra{x}\otimes\ket{H}\bra{H}+\ket{x+1}\bra{x}\otimes\ket{V}\bra{V}\right)$ by a beam displacer (BD) whose optical axis is cut so that the photons in $\ket{V}$ are directly transmitted and those in $\ket{H}$ undergo a lateral displacement into a neighboring spatial mode, respectively. The loss operator $M$ is implemented by a sandwich-type HWP (at $22.5^\circ$)-PPBS-HWP (at $22.5^\circ$) setup~[16]. Here, PPBS is an abbreviation for partially polarizing beamsplitter and the transmissivities of PPBS are $(T_\text{H},T_\text{V})=(1,1-p)$ for horizontally and vertically polarized photons, respectively.

We construct the raw probability distribution of the walker $P_\text{R}$ at time $t$ by dividing the number of coincidence measurements at APDs using the total number of photon pairs, after correcting for the relative efficiencies of different APDs. The raw probability is then converted into the corrected probability $P_\text{C}(x,t)=\gamma^{2t}P_\text{R}(x,t)$, which is obtained by multiplying the correction factor $\gamma$ for the corresponding step $t$ and represents the probability corresponding to parity-time ($\mathcal{PT}$)-symmetric quantum walks (QWs) governed by $\tilde{U}'$. Whereas, the normalized probability $P_\text{N}(x,t)$ is defined as $P_\text{R}(x,t)/\sum_x P_\text{R}(x,t)$.

\section{$\mathcal{PT}$ symmetry of non-unitary QWs governed by $\tilde{U}'$}
In this section, we discuss $\mathcal{PT}$ symmetry of two-step QWs governed by the non-unitary Floquet operator $\tilde{U}'=F\tilde{M}G$, where $F$, $G$, and $\tilde{M}$ are defined in the main text. $\tilde{U}'$ has $\mathcal{PT}$ symmetry, so long as $\theta_{1,2}(x)=\theta_{1,2}(N-x)$. Under the periodic boundary condition, the symmetry operator is $\mathcal{PT}=\sum_{x\in \mathbb{L}}|x\rangle\langle N-x|\otimes \sigma_z\mathcal{K}$, with $\mathcal{PT}\tilde{U}'\left(\mathcal{PT}\right)^{-1}=\tilde{U}'^{-1}$, where $\mathcal{K}$ is complex conjugation.

Compared to the $\mathcal{PT}$-symmetric QW in Ref.~[17], where both interfaces are involved in the QW dynamics, the configuration here facilitates the study of topological edge states in the $\mathcal{PT}$-symmetric setting, as only edge states localized near $x=0$ are relevant for a walker initialized at $x=0$, when $N$ is much larger than the number of time steps of the QW dynamics. 

We now focus on the homogeneous case with $\theta^{\text{L}}_{1,2}=\theta^{\text{R}}_{1,2}=\theta_{1,2}$, which allows us to write $\tilde{U}'$ in momentum space
\begin{align}
\tilde{U}' &= d_0\one_c-id_1\sigma_x-id_2\sigma_y-id_3\sigma_z,\\
d_0 &= \alpha\left(\cos2k\cos\theta_1\cos\theta_2-\sin\theta_1\sin\theta_2\right),\\
d_1 &= i\beta,\\
d_2 &= \alpha\left(\cos2k\cos\theta_2\sin\theta_1+\cos\theta_1\sin\theta_2\right),\\
d_3 &=-\alpha\sin2k\cos\theta_2,\\
d^2_0&+d^2_1+d^2_2+d^2_3=\alpha^2-\beta^2=1,
\label{eqn:U_FM_hatG}
\end{align}
where $\alpha=\gamma(1+\sqrt{1-p})/2$, $\beta=\gamma(1-\sqrt{1-p})/2$ and $d_i$ ($i=0,1,2,3$) are momentum dependent, $\sigma_{x,y,z}$ are the Pauli matrices, and $\one_c$ is a two-by-two identity matrix.

The eigenvalues of $\tilde{U}'$ are given by $\lambda_{\pm}=d_0\mp i\sqrt{1-d_0^2}$, where $\pm$ are band indices. Note that $\lambda_{+}\lambda_-=1$, which is guaranteed by $\mathcal{PT}$ symmetry of the Floquet operator $\tilde{U}'$.
As we define the effective Hamiltonian through $\tilde{U}'=\exp(-i H_{\rm eff})$, the quasienergy spectrum of $H_{\rm eff}$ is given by $\epsilon_{\pm}=i\ln(\lambda_{\pm})$. Apparently, when $d_0^2<1$ for all $k$, the quasienergy spectrum is entirely real. In this case, the system is in the $\mathcal{PT}$-symmetry-unbroken regime. In contrast, when $d_0^2>1$ is satisfied for a certain range of momenta $k$, the corresponding quasienergies in that range become complex. In this case, the system is in the $\mathcal{PT}$-symmetry-broken regime. The transition between the above two scenarios, the so-called exceptional point, occurs when $d_0^2=1$ is satisfied at some discrete momenta while $d_0^2< 1$ otherwise.
At these momenta, the quasienergy band gap closes at $\epsilon=0$ (with $\epsilon_+=\epsilon_-=0$) or $\epsilon=\pi$ (with $\epsilon_+=\epsilon_-=\pi$).

We further divide the $\mathcal{PT}$-symmetry-broken regime into the partially-broken and the completely-broken regimes, where complex quasienergies occur for part of or the whole first Brillioun zone, respectively. The important difference between the $\mathcal{PT}$-symmetry-partially-broken and completely-broken cases is that the latter does not have quasienergy band gap closing, i.e., $(\epsilon_{\pm}\neq 0, \pi)$ for any $k$ in the symmetry completely-broken regime. In the left two columns of Fig.~\ref{fig:figS1}, we plot quasienergies $\epsilon_{\pm}$ and eigenvalues $\lambda_{\pm}$ for the different scenarios above.

We experimentally confirm $\mathcal{PT}$ symmetry of $\tilde{U}'$ by analyzing homogeneous QWs for both the $\mathcal{PT}$-unbroken and broken states. We start with a homogeneous QW in $\mathcal{PT}$-symmetry-unbroken regime, with the coin parameters $(\theta^\text{L,R}_1,\theta^\text{L,R}_2)=(-\pi/4,3\pi/4-3\xi)$. We fix the parameter $\xi=0.1113$ in our experiment.
As illustrated in Fig.~\ref{fig:figS1}(a), all quasienergies are real and gaps are open for all momenta. Correspondingly, eigenvalues of $\tilde{U}'$ all lie on a unit circle in the complex plane. The measured corrected probability distribution is ballistic, which agrees well with numerical simulations and is similar to that of a standard unitary QW.

We then change the coin parameters to $(\theta_1,\theta_2)=(-4\pi/9,5\pi/9+\xi)$. The resulting QW is at the exceptional point. As illustrated in Fig.~\ref{fig:figS1}(b), in this case all the quasienergies are still real, but the quasienergy gap closes at $\epsilon=0$. The measured corrected probability distribution is different from that of the standard unitary QW with a squeezed profile.

When we change the coin parameters to $(\theta_1,\theta_2)=(-17\pi/36,19\pi/36+\xi/2)$, the system enters the $\mathcal{PT}$-symmetry-partially-broken regime. As illustrated in Fig.~\ref{fig:figS1}(c), the quasienergies become complex. Meanwhile, some eignevalues $\lambda_{\pm}$ deviate from the unit circle with $\lambda_{\pm}>0$, which corresponds to $\text{Re} (\epsilon_{\pm})=0$. The corrected probability distribution is Gaussian-like, which is completely different from that in the $\mathcal{PT}$-symmetry-unbroken regime.

When we change the coin parameters to $(\theta_1,\theta_2)=(-\pi/2-3\xi/8,-\pi/2-3\xi/8)$, the system enters the $\mathcal{PT}$-symmetry-completely-broken regime. As illustrated in Fig.~\ref{fig:figS1}(d), all quasienergies become imaginary with $\text{Re}(\epsilon_{\pm})=0$, while the quasienergy spectrum is fully gapped as $\epsilon_{\pm}\neq 0, \pi$. Meanwhile, all eignevalues $\lambda_{\pm}$ deviate from the unit circle and lie on the real axis with $\lambda_{\pm}>0$, which corresponds to $\text{Re} (\epsilon_{\pm})=0$. Again, the corrected probability distribution is Gaussian-like.

\begin{figure*}
\includegraphics[width=\textwidth]{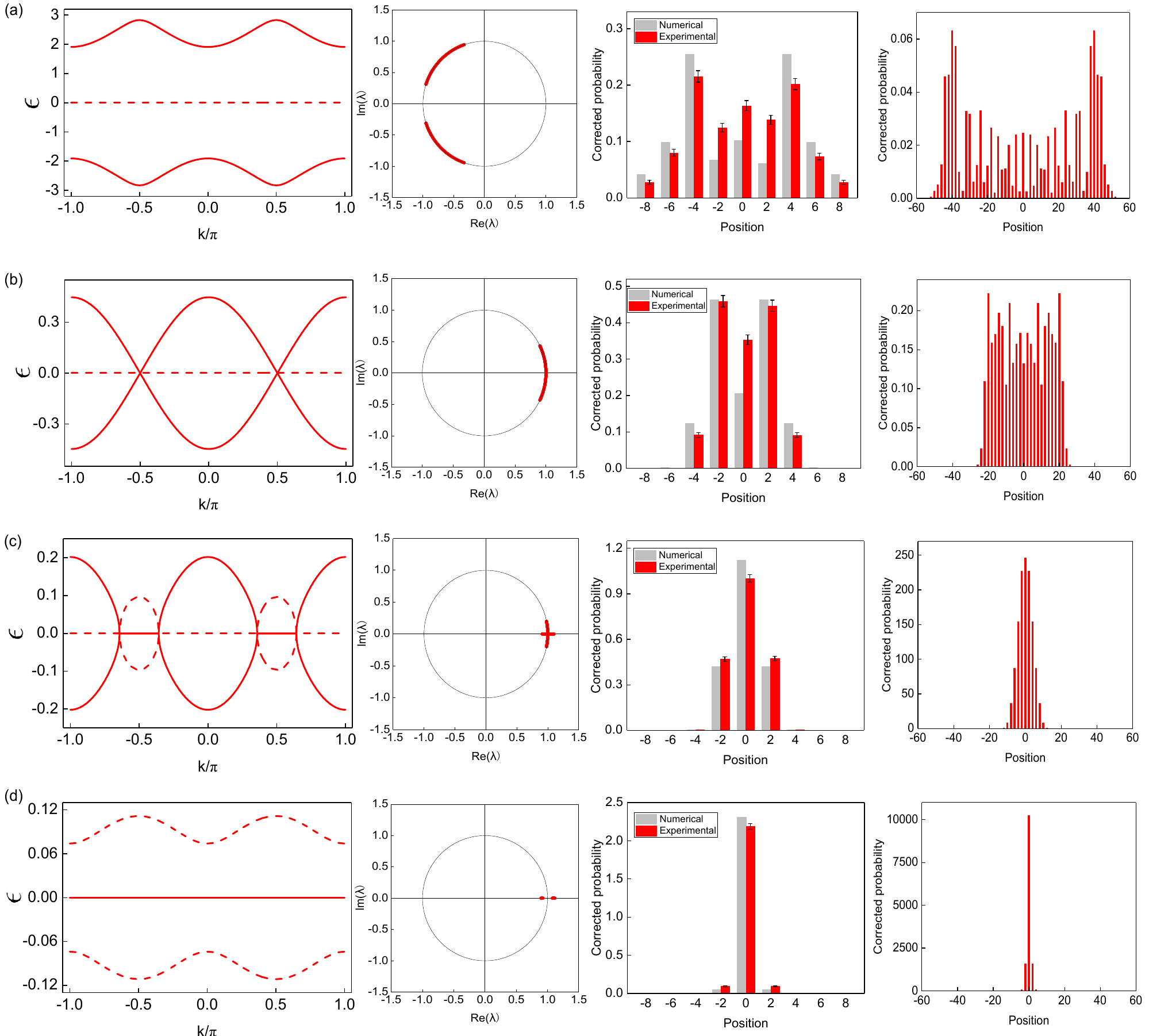}
\caption{Experimental $\mathcal{PT}$-symmetric homogeneous QW with the initial state $\ket{0}\otimes(\ket{+}+i\ket{-})/\sqrt{2}$. (a) QWs with  $(\theta_1,\theta_2)=(-\pi/4,3\pi/4-3\xi)$ in the unbroken $\mathcal{PT}$-symmetric phase. (b) QW with
$(\theta_1,\theta_2)=(-4\pi/9,5\pi/9+\xi)$ at the exceptional point. (c) QW with $(\theta_1,\theta_2)=(-17\pi/36,19\pi/36+\xi/2)$ in the partially broken $\mathcal{PT}$-symmetric phase. (d) QW with $(\theta_1,\theta_2)=(-\pi/2-3\xi/8,-\pi/2-3\xi/8)$ in the completely broken $\mathcal{PT}$-symmetric phase. The first column: the quasienergy as a function of quasimomentum where the solid (dashed) curves represent the real (imaginary)
part of quasienergy. The second column: analytical results of the eigenvalues of the time-evolution operator in the complex plane. The third column: comparison between the measured (red bars) and the predicted (grey bars) probabilities after the seventh step with different coin parameters.  The fourth column: the predicted probabilities after fifty steps with different coin parameters. Experimental errors are due to photon-counting statistics and represent the corresponding standard deviations.}
\label{fig:figS1}
\end{figure*}

\begin{figure*}
\includegraphics[width=0.75\textwidth]{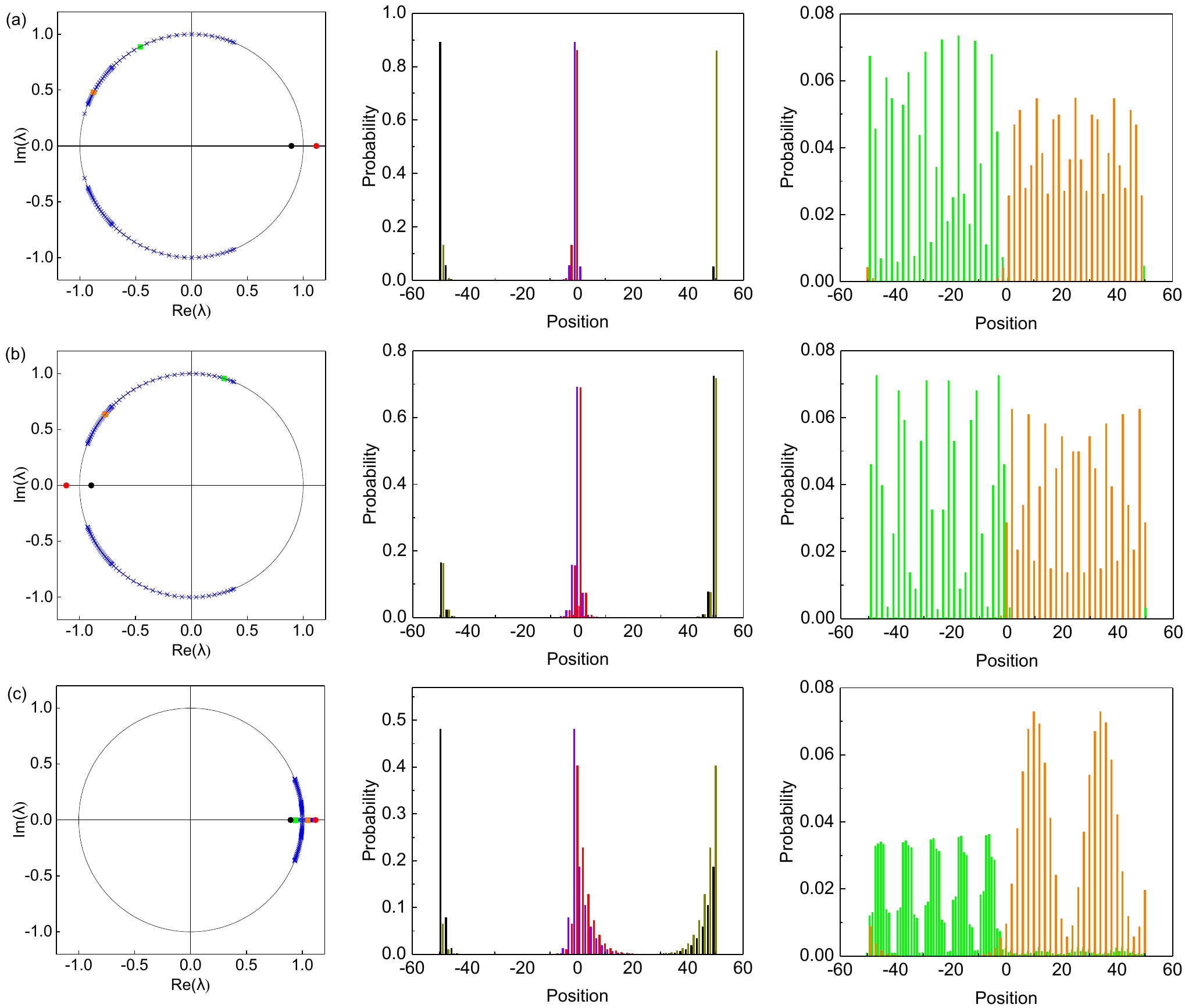}
\caption{Numerical evidence for the bulk-boundary correspondence. We numerically diagonalize Floquet operators for inhomogeneous QWs and examine the relation between the difference in bulk topological numbers and the number of localized edge states. The left column demonstrates the eigenvalue spectra $\lambda$ for different cases on the complex plane. The central column shows the spatial probability distribution of localized topological edge states. The right column shows the spatial probability distribution of extended bulk states. (a) Coin parameters are the same as those in Fig.~2(b) of the main text. (b) Coin parameters are $(\theta_1^{\text{L}},\theta_2^{\text{L}})=(\pi/16,5\pi/16)$, $(\theta_1^{\text{R}},\theta_2^{\text{R}})=(7\pi/16,11\pi/16)$. (c) Coin parameters are the same as those in Fig.~3(c) of the main text.
}
\label{fig:figS2}
\end{figure*}

\begin{figure*}
\includegraphics[width=0.75\textwidth]{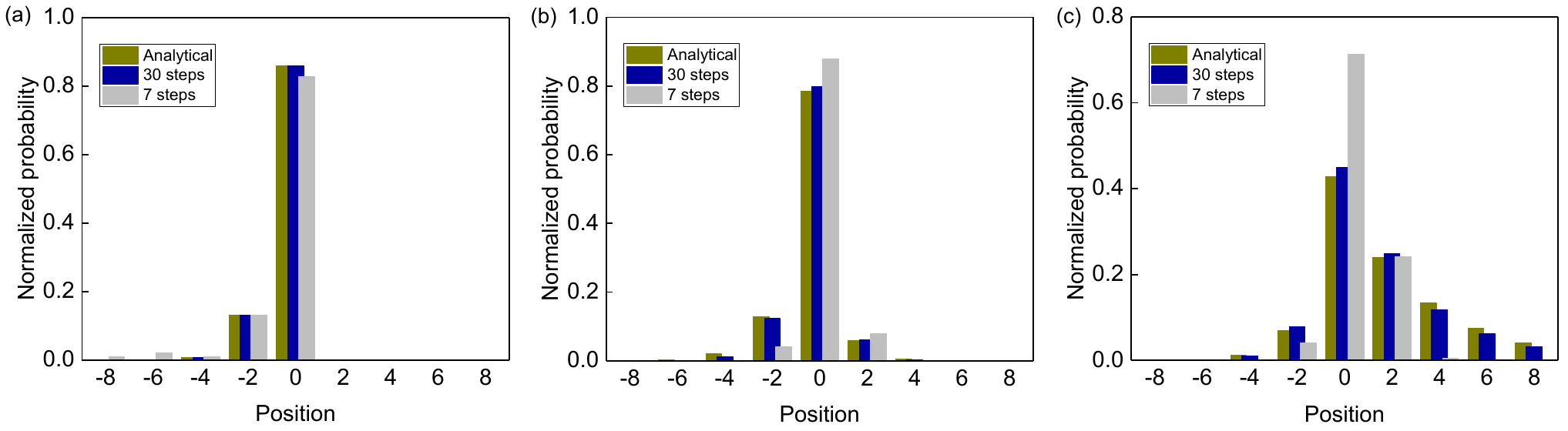}
\caption{Confirming edge-state wave functions with numerical simulations. We compare normalized probability distributions calculated from analytical edge-state wave functions (cyan) with those from numerical simulations after seven (grey) and thirty (blue) time steps. (a) Topological edge states between $\mathcal{PT}$-symmetry-unbroken bulks, with the same coin parameters and initial states as those of Fig.~2(b) in the main text. (b) Topological edge states between a $\mathcal{PT}$-symmetry-unbroken bulk and a broken one. Coin parameters and initial states as the same as those of Fig.~3(b) in the main text. (c) Topological edge states between $\mathcal{PT}$-symmetry-broken bulks. Coin parameters and initial states as the same as those of Fig.~3(c) in the main text.
}
\label{fig:figS3}
\end{figure*}


\begin{figure}
\includegraphics[width=0.75\textwidth]{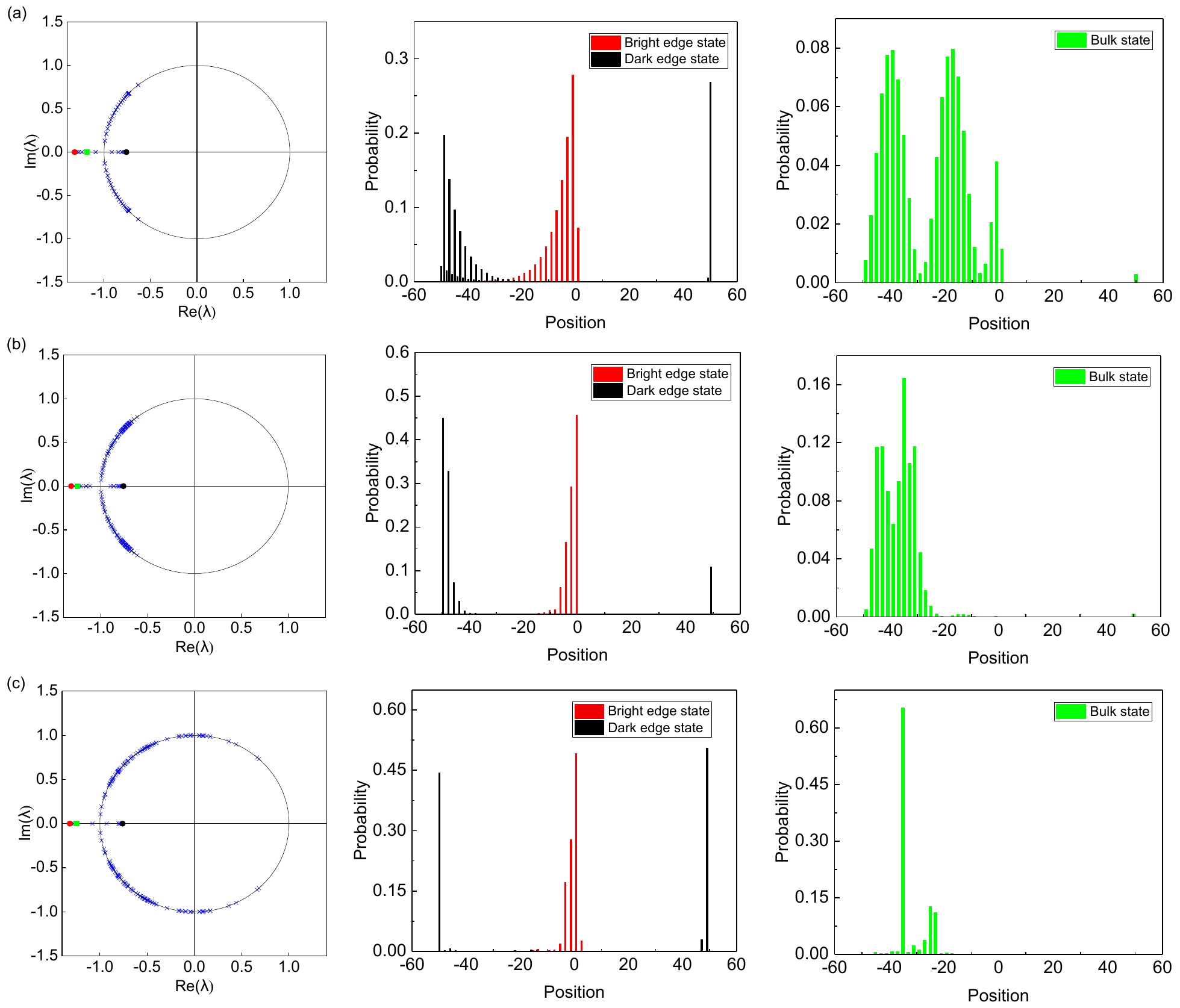}
\caption{The influence of disorder on edge states and bulk states. Left column: the eigenvalue spectra $\lambda$ for three cases [(a) no disorder, (b) weak disorder, and (c) strong disorder] on the complex plane. Central column: the spatial probability distribution of localized topological edge states. Right column: the spatial probability distribution of bulk states.}
\label{fig:figS5}
\end{figure}

\section{Global Berry phase of $\mathcal{PT}$-symmetric QWs}
In this section, we discuss the definition of topological invariants for $\mathcal{PT}$-symmetric non-unitary QWs $\tilde{U}'$. For the convenience of calculation, we apply a unitary transformation to $\tilde{U}'$
\begin{equation}
W' = V\tilde{U}'V^\dag=d_0\one_c-i(-d_3)\sigma_x-id_2\sigma_y-id_1\sigma_z,
\end{equation}
where $V=e^{i\frac{\pi/2}{2}\sigma_y}$. Topological properties of $\tilde{U}'$ is not changed under the unitary transformation. We will show that the winding number $\nu'$ is defined through the global Berry phase as $\nu'=\varphi_\text{B}/2\pi$. Here, $\varphi_\text{B}=\varphi_{Z+}+\varphi_{Z-}$, with the generalized Zak phases for the $j$th band ($j=\pm$)
\begin{align}
\varphi_{Z\pm} &=-i\oint \text{d}k\frac{\langle\chi_{\pm}|\frac{\text{d}}{\text{d}k}|\psi_{\pm}\rangle}{\langle\chi_{\pm}|\psi_{\pm}\rangle}.
\end{align}
Here, the integral is over the first Brillioun zone and $\langle\chi_{j}|$ and $|\psi_{j}\rangle$ are respectively the left and right eigenstates of $W'$, defined through $W'^{\dag}|\chi_j\rangle=\lambda_j^* |\chi_j\rangle$ and $W'|\psi_j\rangle=\lambda_j|\psi_j\rangle$, respectively.

In the following, let us first evaluate the Berry connection
\begin{align}
A_\pm=-i\frac{\bra{\chi_{\pm}}\frac{\text{d}}{\text{d}k}\ket{\psi_{\pm}}}{\langle\chi_{\pm}|\psi_{\pm}\rangle},
\end{align}
which critically depends on whether $d_0^2$ is greater than $1$ or not. As different momenta are decoupled, we will examine the Berry connection case by case.

{\it Case I: the momentum region with $d^2_0<1$:---}
In the momentum regime with $d^2_0<1$, we have $d_1^2+d_2^2+d_3^2>0$, and the right and the left eigenvectors of $W'$ are
\begin{align}
\ket{\psi_{\pm}}&=\frac{1}{\sqrt{2\cos2\Omega}}(\pm e^{\pm i\Omega},e^{+i\vartheta}e^{\mp i\Omega})^\text{T},\\
\bra{\chi_{\pm}}&=\frac{1}{\sqrt{2\cos2\Omega}}(\pm e^{\pm i\Omega},e^{-i\vartheta}e^{\mp i\Omega}).
\end{align}
Here $\vartheta$ and $\Omega$ are respectively defined through $-d_3+i d_2 = d e^{i\vartheta}$ and $\sin2\Omega = -id_1/d$, with $d^2=d_2^2+d_3^2$. As $d_1/d\in(0,1)$, we set $2\Omega\in(0,\pi/2)$ and $\cos2\Omega>0$. Notice that the orthonormal conditions $\left(\langle\chi_{\pm}|\psi_{\pm}\rangle = 1, \langle\chi_{\pm}|\psi_{\mp}\rangle = 0\right)$ are always satisfied in this region.
We then have
\begin{equation}
\frac{\text{d}}{\text{d}k}\ket{\psi_{\pm}}=\frac{1}{(2\cos2\Omega)^{3/2}}(i2e^{\mp i\Omega}\Omega',ie^{i\vartheta}e^{\mp i\Omega}(2\cos2\Omega)\vartheta'\mp i2e^{i\vartheta}e^{\pm i\Omega}\Omega')^\text{T},
\end{equation}
where $\Omega'=\text{d}\Omega/\text{d}k$ and $\vartheta'=\text{d}\vartheta/\text{d}k$. It is then straightforward to derive $A_\pm=\frac{1}{2}\vartheta'\pm\frac{i}{2}\vartheta'\tan2\Omega$, and $A_++A_-=\vartheta'$.

{\it Case II: the momentum region with $d^2_0>1$:---}
In the momentum regime with $d^2_0>1$, we have $d_1^2+d_2^2+d_3^2<0$, and the right and the left eigenvectors of $W'$ are
\begin{align}
\ket{\psi_{\pm}} &=\frac{1}{\sqrt{\mp2\sinh2\Xi}}(ie^{\pm\Xi},e^{+i\vartheta}e^{\mp\Xi})^\text{T},\\
\bra{\chi_{\pm}} &=\frac{1}{\sqrt{\mp2\sinh2\Xi}}(ie^{\pm\Xi},e^{-i\vartheta}e^{\mp\Xi}),
\end{align}
where $\cosh2\Xi = -id_1/d$, with $\Xi\in(0,\infty)$.
We then have
\begin{equation}
\frac{\text{d}}{\text{d}k}|\psi_{\pm}\rangle=\frac{1}{(\mp2\sinh2\Xi)^{3/2}}(\pm i2\Xi'e^{\mp\Xi},\mp i2\vartheta'\sinh2\Xi e^{i\vartheta}e^{\mp \Xi}\pm2\Xi'e^{i\vartheta}e^{\pm\Xi})^\text{T},
\end{equation}
where $\Xi'=\text{d}\Xi/\text{d}k$. The Berry connection is then $A_\pm=\pm\frac{e^{\mp 2\Xi}}{2\sinh2\Xi}\vartheta'$. Again, we have  $A_++A_-=\vartheta'$.

{\it Case III: at discrete momenta with $d_0^2=1$:---}
In this case, $d_1^2+d_2^2+d_3^2=0$. The right and the left eigenvectors of $W'$ are
\begin{align}
\ket{\psi_{\pm}} =&\frac{1}{\sqrt{2}}(i,e^{+i\theta})^\text{T},\\
\bra{\chi_{\pm}} =&\frac{1}{\sqrt{2}}(i,e^{-i\theta}).
\end{align}
As $\langle\chi_{\pm}|\psi_{\pm}\rangle = \langle\chi_{\pm}|\psi_{\mp}\rangle = 0$ and $\langle\chi_{\pm}|\chi_{\pm}\rangle =\langle\psi_{\pm}|\psi_{\pm}\rangle =1$, the denominator in the Berry connection $\langle\chi_{\pm}|\psi_{\pm}\rangle$ vanishes, giving rise to diverging $A_{\pm}$ at these momenta. However, a closer examination reveals that the divergence in $A_+$ and $A_-$ cancels out in their summation $A_++A_-$, giving rise to a well-defined ``global Berry connection''. For example, if the condition $d^2_0=1$ is approached in parameter space from the side with $d^2_0<1$, we have $A_\pm=\frac{1}{2}\vartheta'\pm\frac{i}{2}\vartheta'\tan2\Omega$. At $d^2_0=1$, $\tan2\Omega\rightarrow \infty$. The imaginary parts of Berry connections $A_\pm$ diverge, however, their summation is still $\vartheta'$ and remains well-defined even at $d^2_0=1$. The situation is similar when the condition $d^2_0=1$ is approached in parameter space from the side with $d^2_0>1$, where $A_\pm$ diverge but their sum is not.
With the above analysis, we see that the global Berry phase is given by $\varphi_\text{B}=\oint \text{d}\vartheta$, regardless of whether the system is $\mathcal{PT}$-symmetry unbroken or broken.

In contrast, at the topological phase boundary, the polar angle $\vartheta$ in the above expressions become ill-defined as $d_2=d_3=0$. This occurs at $k=0,\pi$ for $d_0(k)=\alpha$, or at $k=\pm\pi/2$ for $d_0(k)=-\alpha$. As such, the global Berry phase can no longer be defined on the topological phase boundary. This conclusion is the same as the Zak phase of a unitary QW.

In previous studies, both generalized Zak phases and generalized winding numbers have been proposed to serve as topological invariants for non-unitary FTPs. The generalized Zak phase, which is only valid in the $\mathcal{PT}$-symmetry-unbroken regime, can be written as $\text{Re}\left(\varphi_{Z-}\right)$. As $\varphi_{Z-}=\varphi_{Z+}^\ast$ in the $\mathcal{PT}$-symmetry-unbroken regime, the generalized Zak phase defined in Refs.~[17,22] is equivalent to the global Berry phase $\varphi_\text{B}$ in the $\mathcal{PT}$-symmetry-unbroken regime.

On the other hand, according to Refs.~[15,16], the generalized winding number for the Floquet operator $\tilde{U}'$ is defined as
\begin{equation}
\nu_1=\frac{1}{2\pi}\oint \text{d}k\left(\bm{n}\times\frac{\partial{\bm{n}}}{\partial{k}}\right)_x,
\label{eqn:windingnumber}
\end{equation}
where the unit vector $\bm{n}$ is a normalized projection of the vector $\bm{d}=(d_1,d_2,d_3)^\text{T}$ in the $y$-$z$ plane. As such, $\bm{n}=(0,d_2/d,d_3/d)^\text{T}$. It is then straightforward to show that $\vartheta'=(\bm{n}\times\frac{\partial{\bm{n}}}{\partial{k}})_x$, such that the generalized winding number is $\nu_1=\varphi_\text{B}/2\pi=\nu'$. Topological invariants defined through the global Berry phase thus unify previous definitions in different contexts.

\section{Topological number and topological edge states}

In this section, we numerically confirm that localized topological edge states at a given boundary are dictated by the difference in topological numbers $(\nu_0,\nu_\pi)$ of the bulks on either side. More specifically, topological number $\nu_0$ ($\nu_\pi$) is associated with the number of topological edge states with $\text{Re}(\epsilon)=0$ [$\text{Re}(\epsilon)=\pi$]. For convenience, we define $\Delta\nu_g=|\nu_g^{\text{L}}-\nu_g^{\text{R}}|$ ($g=0,\pi$), where $\nu_g^{\text{L}}$ ($\nu_g^{\text{R}}$) is the topological number in the left (right) region.

We numerically diagonalize Floquet operators of inhomogeneous QWs governed by $\tilde{U}'$ with $N=50$. As illustrated in Fig.~\ref{fig:figS2}(a), when the difference in topological numbers is $(\Delta\nu_0,\Delta\nu_\pi)=(2,0)$, a pair of degenerate edge states, on odd and even sites respectively, exist at a given boundary ($x=0$ or $x=50$) with $\text{Re}(\epsilon)=0$ [$\text{Re}(\lambda)>0$]. We note that both regions belong to the $\mathcal{PT}$-symmetry-unbroken regime.
In this case, edge states with $\lambda>1$ (red) appear near $x=0$. In contrast, for bulk states with $|\lambda|=1$, their spatial distributions are extended (green and orange).

In Fig.~\ref{fig:figS2}(b), both regions are in the $\mathcal{PT}$-symmetry-unbroken regime, and the difference in topological numbers is $(\Delta\nu_0,\Delta\nu_\pi)=(0,2)$. A pair of degenerate edge states, on odd and even sites respectively, exist at a given boundary with $\text{Re}(\epsilon)=\pi$ [$\text{Re}(\lambda)<0$]. In this case, edge states with $\lambda<-1$ (red) appear near $x=0$. In contrast, for bulk states with $|\lambda|=1$, their spatial distributions are extended (green and orange).

In Fig.~\ref{fig:figS2}(c), we consider the case where both left and right regions are in the $\mathcal{PT}$-symmetry-broken regime. The difference in topological numbers is $(\Delta\nu_0,\Delta\nu_\pi)=(2,0)$. In this case, while $\mathcal{PT}$-symmetry-broken bulk states exist with $|\lambda|\neq 1$, localized edge states can still be identified as their eigenvalues $\lambda$ deviate most from the unit circle. In terms of quasienergy, topological edge states in this case possess quasienergies with the largest imaginary parts. As shown in the central and right columns, localized edge states and extended bulk states are also differentiated by their distinct spatial probability distributions.

To summarize, we have numerically confirmed that the number of localized edge states is governed by $\Delta\nu_0$ and $\Delta\nu_\pi$. We have further checked (not shown) that such a relation holds when both topological numbers are different. Such a bulk-boundary correspondence exists even when one or both of the bulks are in the $\mathcal{PT}$-symmetry-broken regime. For all the cases shown in Fig.~\ref{fig:figS2}, edge states exist near $x=0$ ($x=50$). We note that their spatial location would be switched when we exchange the coin parameters of the left and right regions.

\section{Edge-state wave functions}
In this section, we solve for the wave function of topological edge states in $\mathcal{PT}$-symmetry non-unitary QWs governed by $\tilde{U}'$. We first consider the unitary QW with $p=0$ ($\gamma=1$), and derive the analytical solution of topological edge states near the boundary $x=0$. We then demonstrate that topological edge states in the non-unitary case ($\gamma\neq 1$) have the same eigen wave functions
as the unitary case. The difference lies in eigenvalues and hence the time evolution, where topological edge states in the non-unitary case acquire factors $\gamma^{\pm t}$, giving rise to the edge states as discussed in the main text.

\subsection{Edge states in the unitary case}
In the unitary case, we write the Floquet operator as $\bar{U}'=FG$,
where $F$ and $G$ are defined in the main text. In the homogeneous case with $\theta^{\text{L}}_{1,2}=\theta^{\text{R}}_{1,2}=\theta_{1,2}$,
eigenvalues and eigenstates of $\bar{U}'$ are given in momentum space as
\begin{align}
\bar{\lambda}_+ =& e^{-iE},~\ket{\bar{\psi}_+}=\begin{bmatrix}-\cos\theta_2\sin 2k+\sin E
\\i(\cos\theta_2\sin\theta_1\cos 2k+\cos\theta_1\sin\theta_2)\end{bmatrix},\nonumber\\
\bar{\lambda}_- =& e^{iE},~\ket{\bar{\psi}_-}=\begin{bmatrix}-\cos\theta_2\sin 2k-\sin E
\\i(\cos\theta_2\sin\theta_1\cos2k+\cos\theta_1\sin\theta_2)\end{bmatrix},\label{eqn:dispersion}
\end{align}
where $\bar{U}'|\bar{\psi}_\pm\rangle=\bar{\lambda}_\pm|\bar{\psi}_\pm\rangle$, and the quasienergy $E$ satisfies $\cos E = \cos\theta_1\cos\theta_2\cos2k-\sin\theta_1\sin\theta_2$.

In the homogeneous case, eigenstates of  $\bar{U}'$ at a given momentum $k$ can be written as
\begin{align}
|\bar{\psi}_\pm(x)\rangle =e^{ikx}\ket{\bar{\psi}_\pm}.
\label{eqn:planewaves}
\end{align}

In the inhomogeneous case with $\theta^{\text{L}}_{1,2}\neq\theta^{\text{R}}_{1,2}$, topological edge states with quasienergies $E^{(0)}=0$ or $E^{(\pi)}=\pi$ emerge near boundaries between different bulk topological phases. Wave functions of topological edge states near the boundary $x=0$ can be constructed from Eq.~(\ref{eqn:planewaves}) by setting $k=-i\kappa_{\text{L}}$ and $k=i\kappa_{\text{R}}$ for the left and right regions, respectively. Note $\text{Re} \left(\kappa_{\text{L}}\right),\text{Re} \left(\kappa_{\text{{R}}}\right)>0$ so that probability distributions of the localized edge states vanish as $|x|\rightarrow \infty$. We further notice that under the two-step QW $\bar{U}'$, wave functions on even sites and odd sites are decoupled.

The considerations above enable us to construct wave functions for topological edge states at the boundary near $x=0$
\begin{align}
|\psi^{\text{o}(\text{e})}(x)\rangle=
\begin{cases}
r^{\text{o}(\text{e})}e^{\kappa_{\text{L}}x}(a^{\text{L}}_{-i\kappa_{\text{L}}},b^{\text{L}}_{-i\kappa_{\text{L}}})^\text{T},~x<0,\\
t^{\text{o}(\text{e})}e^{-\kappa_{\text{R}}x}(a^{\text{R}}_{i\kappa_{\text{R}}},b^{\text{R}}_{i\kappa_{\text{R}}})^\text{T},~x\geq 0,
\end{cases}
\label{eqn:edgeansatz}
\end{align}
where $|\psi^{\text{o}(\text{e})}(x)\rangle$ is the edge-state wave function on odd (even) sites, and
$r^{\text{o}(\text{e})}$ and $t^{\text{o} (\text{e})}$ are the corresponding coefficients. We also have $\begin{bmatrix}a^{\xi}_{k}\\b^{\xi}_{k}\end{bmatrix}=\begin{bmatrix}-\cos\theta^{\xi}_2\sin 2k
\\i(\cos\theta^\xi_2\sin\theta^\xi_1\cos 2k+\cos\theta^\xi_1\sin\theta^\xi_2)\end{bmatrix}$ ($\xi=\text{L},\text{R}$), which, according to Eq.~(\ref{eqn:dispersion}), denotes coin states of edge-state wave functions with $E^{(0)}$ or $E^{(\pi)}$. We will show in the following that all the coefficients above have analytical forms.

From the dispersion relations, we first establish expressions for the spatial decay rates $\kappa_{\xi}$ ($\xi=\text{L},\text{R}$)
\begin{align}
\cosh 2\kappa_\xi=\frac{\cos E^{(0,\pi)}+\sin\theta^\xi_1\sin\theta^\xi_2}{\cos\theta^\xi_1\cos\theta^\xi_2}.
\label{eqn:decayrates}
\end{align}
From Eq.~(\ref{eqn:decayrates}), it is immediately clear that spatial decay rates to the left (right) of the boundary are determined by the corresponding coin parameters $\theta_{1,2}^{\text{L}}$ ($\theta_{1,2}^{\text{R}}$), as well as the the quasienergy of the edge state $E^{(0,\pi)}$. The edge-state spatial wave function is therefore typically asymmetric with respect to the boundary.

On the other hand, $\bar{U}'$ has chiral symmetry with the symmetry operator $\Gamma=\sum_x|x\rangle\langle x|\otimes\sigma_x$ and $\Gamma \bar{U}'\Gamma=\bar{U}'^{-1}$. Edge states are therefore eigenstates of the chiral operator, such that $\frac{a_{-i\kappa_{\text{L}}}}{b_{-i\kappa_{\text{L}}}}=\frac{a_{i\kappa_{\text{R}}}}{b_{i\kappa_{\text{R}}}}=\pm1$. Equivalently, the edge states are either in $|+\rangle$ or $|-\rangle$.
Combining the expressions of $a_k$ and $b_k$, we derive conditions for the coin states. Specifically, edge states with quasienergy $E^{(0,\pi)}$ are in $|+\rangle$ when
\begin{align}
\begin{cases}
\sinh 2\kappa_{\text{L}}=\frac{\cos E^{(0,\pi)}\sin\theta^{\text{L}}_1+\sin\theta^{\text{L}}_2}{\cos\theta^{\text{L}}_1\cos\theta^{\text{L}}_2},\\
\sinh 2\kappa_{\text{R}}=-\frac{\cos E^{(0,\pi)}\sin\theta^{\text{R}}_1+\sin\theta^{\text{R}}_2}{\cos\theta^{\text{R}}_1\cos\theta^{\text{R}}_2}.
\end{cases}
\label{eqn:cond1}
\end{align}
And edge states are in $|-\rangle$ when
\begin{align}
\begin{cases}
\sinh 2\kappa_{\text{L}}=-\frac{\cos E^{(0,\pi)}\sin\theta^{\text{L}}_1+\sin\theta^{\text{L}}_2}{\cos\theta^{\text{L}}_1\cos\theta^{\text{L}}_2},\\
\sinh 2\kappa_{\text{R}}=\frac{\cos E^{(0,\pi)}\sin\theta^{\text{R}}_1+\sin\theta^{\text{R}}_2}{\cos\theta^{\text{R}}_1\cos\theta^{\text{R}}_2}.
\end{cases}\label{eqn:cond2}
\end{align}

Finally, we show how to solve for the coefficients $r^{\text{o}(\text{e})}$ and $t^{\text{o}(\text{e})}$ by considering the Floquet operator $\bar{U}''=GF$. As discussed in Ref.~[37], $\bar{U}''$ has chiral symmetry and support topological edge state at boundaries between regions with different topological numbers. Further, we notice that  $\bar{U}'(\theta^{\xi}_1,\theta^\xi_2)=\bar{U}''(\theta^\xi_2,\theta^\xi_1)$, and that Eqs.~(\ref{eqn:cond1}) and (\ref{eqn:cond2}) acquire different signs on exchanging $\theta^\xi_1$ and $\theta^\xi_2$. We then establish that edge states under $\bar{U}'$ and $\bar{U}''$ have the same coin states at $E^{(0)}$  ($\bar{\lambda}=1$), and they have opposite coin states at $E^{(\pi)}$ ($\bar{\lambda}=-1$).

On the other hand, for topological edge states $|\psi^{\text{o}(\text{e})}\rangle$ satisfying $\bar{U}'|\psi^{\text{o}(\text{e})}\rangle=\pm|\psi^{\text{o}(\text{e})}\rangle$, we have $\bar{U}''G|\psi^{\text{o}(\text{e})}\rangle=\pm G|\psi^{\text{o}(\text{e})}\rangle$. Therefore, $G|\psi^{\text{o}(\text{e})}\rangle$ is the edge state of $\bar{U}''$ with eigenvalues $\bar{\lambda}=\pm 1$. By matching coin states of $|\psi^{\text{o}(\text{e})}\rangle$ and $G|\psi^{\text{o}(\text{e})}\rangle$ according to Eqs.~(\ref{eqn:dispersion}), (\ref{eqn:cond1}), and (\ref{eqn:cond2}), we derive the ratio $r^{\text{o}(\text{e})}/t^{\text{o}(\text{e})}$. Combining the normalization condition $\langle\psi^{\text{o}(\text{e})}|\psi^{\text{o}(\text{e})}\rangle=1$, we can solve for analytical expressions for the coefficients $r^{\text{o}(\text{e})}$, and $t^{\text{o}(\text{e})}$.

As a concrete example, we consider the case
$(\theta^L_1,\theta^L_2)=(\pi/16,5\pi/16)$ and $(\theta^R_1,\theta^R_2)=(-9\pi/16,-5\pi/16)$, and derive the analytical wave function of the topological edge state with quasienergy $E^{(0)}$ on odd sites. According to Eqs.~(\ref{eqn:cond1}) and (\ref{eqn:cond2}), coin states of $\ket{\psi^o}$ and $G\ket{\psi^o}$ are both $|+\rangle$. We then have
\begin{equation}
\begin{cases}
\frac{\sqrt{B}t^o}{\sqrt{A}r^o}=\frac{(\cos\frac{\theta^L_1}{2}+\sin\frac{\theta^L_1}{2})(\cos\frac{\theta^R_2}{2}+\sin\frac{\theta^R_2}{2})}
{(\cos\frac{\theta^R_1}{2}-\sin\frac{\theta^R_1}{2})(\cos\frac{\theta^R_2}{2}-\sin\frac{\theta^R_2}{2})}
:=\tan\alpha,\\
\frac{(\sqrt{A}r^o)^2}{1-A^2}+\frac{(\sqrt{B}t^o)^2}{1-B^2}=\frac{1}{2},
\end{cases}
\Rightarrow
\begin{cases}
r^o=\sqrt{\frac{1-A^2}{2A}}\cos\alpha',\\
t^o=\sqrt{\frac{1-B^2}{2B}}\sin\alpha',
\end{cases}
\end{equation}
where $A=e^{-2\kappa_L}$, $B=e^{-2\kappa_R}$, $\alpha'=\arctan(\sqrt{\frac{1-A^2}{1-B^2}}\tan\alpha)$, and $\alpha\in(-\pi/2,\pi/2)$. The analytical solutions of the edge-state wave function agree well with numerical results.

\subsection{Edge-state wave function for non-unitary QWs}

Consider topological edge states of $\bar{U}'$, with $\bar{U}'|\psi^{\text{o}(\text{e})}\rangle=\bar{\lambda} |\psi^{\text{o}(\text{e})}\rangle$ ($\bar{\lambda}=\pm 1$). Applying the non-unitary Floquet operator, we have
\begin{align}
\tilde{U}'|\psi^{\text{o}(\text{e})}\rangle=F\gamma MG|\psi^{\text{o}(\text{e})}\rangle=
\begin{cases}
\gamma\ket{\psi^{\text{o}(\text{e})}},~\mathrm{if}~\bar{\lambda}=1~\mathrm{and}~\ket{\psi^{\text{o}(\text{e})}}\text{ in coin state }\ket{+},\\
\frac{1}{\gamma}\ket{\psi^{\text{o}(\text{e})}},~\mathrm{if}~\bar{\lambda}=1~\mathrm{and}~\ket{\psi^{\text{o}(\text{e})}}\text{ in coin state }\ket{-},\\
-\gamma\ket{\psi^{\text{o}(\text{e})}},~\mathrm{if}~\bar{\lambda}=-1~\mathrm{and}~\ket{\psi^{\text{o}(\text{e})}}\text{ in coin state }\ket{-},\\
-\frac{1}{\gamma}\ket{\psi^{\text{o}(\text{e})}},~\mathrm{if}~\bar{\lambda}=-1~\mathrm{and}~\ket{\psi^{\text{o}(\text{e})}}\text{ in coin state }\ket{+}.
\end{cases}
\end{align}

Therefore, localized edge state of $\bar{U}'$ are also eigenstates of $\tilde{U}'$, with eigenvalues being $\pm\gamma$ or $\pm 1/\gamma$ in the non-unitary case. The corresponding quasiernergies satisfy $\text{Re}(\epsilon)=0,\pi$, which are exactly the conditions for topological edge states as required by pseudo-anti-unitarity of $\tilde{U}'$~[31]. We therefore conclude that topological edge states under $\tilde{U}'$ have the same spatial and coin-state wave functions as those in unitary case. The difference lies in the quasienergies and hence the time evolution.


Due to $\mathcal{PT}$ symmetry of $\tilde{U}'$, eigenstates with the eigenvalues $\lambda$ and $\lambda^{-1}$ must appear in pairs. This implies that edge states must also appear in pairs. From Eqs.~(\ref{eqn:cond1}) and (\ref{eqn:cond2}), we see that at a given boundary, edges states associated with the same topological number ($\nu_0$ or $\nu_{\pi}$) have the same coin states and are of the same type. In fact, they only differ by the occupation of odd or even sites. Thus, edge states associated with the same topological number are two-fold degenerate.

Finally, we confirm the analytical edge-state wave functions derived above by comparing the normalized probability distributions from the analytical solution and from numerical simulations of QW dynamics governed by $\tilde{U}'$. In Fig.~\ref{fig:figS3}, we see that as the time steps of the numerical simulation increase, the resulting normalized probability approaches that of the analytical solution. Apparently, it takes more time steps for the QW dynamics to converge to the edge-state distribution when at least one of the bulks is $\mathcal{PT}$-symmetry broken.

\end{widetext}

\end{document}